\documentclass[%
 reprint,
 amsmath,amssymb,
 aps,
prb,
]{revtex4-1}

\usepackage{graphicx}
\usepackage{dcolumn}
\usepackage{bm}
\usepackage{cleveref }

\begin{document}

\preprint{APS/123-QED}

\title{What is the Brillouin Zone of an Anisotropic Photonic Crystal?}

\author{P. Sivarajah, A. A. Maznev, B.K. Ofori-Okai}
\author{K. A. Nelson}%
 \email{Corresponding author; e-mail: kanelson@mit.edu}
\affiliation{%
 Department of Chemistry,\\ MIT, 77 Massachusetts Avenue, Cambridge, MA 02139, USA\\
}




\date{\today}

\begin{abstract}
The concept of the Brillouin zone (BZ) in relation to a photonic crystal fabricated in an optically anisotropic material is explored both experimentally and theoretically. In experiment, we used femtosecond laser pulses to excite THz polaritons and image their propagation in lithium niobate and lithium tantalate photonic crystal (PhC) slabs. We directly measured the dispersion relation inside PhCs and observed that the lowest bandgap expected to form at the BZ boundary forms inside the BZ in the anisotropic lithium niobate PhC. Our analysis shows that in an anisotropic material the BZ -- defined as the Wigner-Seitz cell in the reciprocal lattice -- is no longer bounded by Bragg planes and thus does not conform to the original definition of the BZ by Brillouin.  We construct an alternative Brillouin zone defined by Bragg planes and show its utility in identifying features of the dispersion bands.  We show that for an anisotropic 2D PhC without dispersion, the Bragg plane BZ can be constructed by applying the Wigner-Seitz method to a stretched or compressed reciprocal lattice. We also show that in the presence of the dispersion in the underlying material or in a slab waveguide, the Bragg planes are generally represented by curved surfaces rather than planes. The concept of constructing a BZ with Bragg planes should prove useful in understanding the formation of dispersion bands in anisotropic PhCs and in selectively tailoring their optical properties. 
\end{abstract}

\maketitle


\section{\label{sec:intro}Introduction}

Photonic crystals (PhCs) have played a major role in both fundamental science and technology since their initial proposition in 1987 \cite{John1987a,Yablonovitch1987a}. In a photonic crystal, the refractive index modulation encountered by the electromagnetic (EM) wave is analogous to the periodic potential encountered by the electron’s wave function in natural crystals. In both cases, eigenmodes of a periodic medium are no longer represented by plane waves characterized by a unique wavevector. Rather, an eigenmode is represented by a Bloch wave \cite{Joannopoulos2008} comprising a linear combination of plane waves (Bloch harmonics). 

In a Bloch wave, the wavevector $\mathbf{k}$  (or crystal momentum in the band theory of solids) is not uniquely defined -- it is equivalent to a wavevector  $\mathbf{k+G}$, where $\mathbf{G}$ is the reciprocal lattice vector. Consequently, it is convenient to introduce a uniquely defined reduced wavevector within a primitive unit cell of the reciprocal lattice.  Just as in the case of a direct lattice, many choices of the primitive cell of the reciprocal lattice are possible; however, the primitive cell universally used is the first Brillouin zone, often referred to simply as the Brillouin zone (BZ). The concept of Brillouin zones was formulated by Brillouin \cite{Brillouin1953} based on Bragg planes, i.e. planes formed by wavevectors of plane waves satisfying the Bragg reflection condition in the limiting case of a small periodic perturbation. The first BZ is defined as the domain enclosed around $\mathbf{k}=0$ without crossing any Bragg planes. In an isotropic medium, Bragg planes are defined by planes which bisect neighboring reciprocal lattice points, and so the BZ can be alternatively defined as a Wigner-Seitz (WS) primitive cell of the reciprocal lattice, i.e. a domain around a given lattice point within which the distance to this lattice point is smaller than to any other lattice point \cite{Ashcroft1976a}. Since for an isotropic medium the two definitions are equivalent, the BZ is often thought of as the WS cell of the reciprocal lattice without reference to Bragg planes \cite{Joannopoulos2008}. 

However, the equivalence of the original definition of the BZ and the WS cell breaks down in PhCs fabricated in a medium with an anisotropic refractive index (AnPhC), which have garnered significant interest with the rapid development of the field. Two of the main advantages of AnPhCs are the ability to tune their dispersion by controlling the relative anisotropy (e.g. using liquid crystals) \cite{Yoshino1999} and their ability to form complete photonic bandgaps by lifting the degeneracy of the EM modes \cite{Li1998b}.

As we will show below, in an anisotropic medium the WS cell of the reciprocal lattice is no longer bounded by Bragg planes. One is then faced with the following choice in defining the BZ: (i) Stick to the WS cell and continue to call it the BZ even though it no longer conforms to the original definition of the BZ. This is the choice commonly made by researchers studying AnPhCs \cite{Alagappan2006,Khromova2008} and it has the advantage of simplicity in constructing the primitive cell. However, one should recognize that this Wigner-Seitz BZ (WSBZ) will not be as useful as the BZ of an isotropic refractive index PhC (IsPhC) in characterizing the dispersion of eigenmodes. In particular, we will show that for AnPhCs the bandgaps between the first and second dispersion bands may no longer form along the boundary of WSBZ. (ii) Try to construct the Brillouin zone based on its original definition, i.e. using Bragg planes. This is the alternative explored in the present work. 

Our study is motivated by an experiment in which we studied the propagation of THz phonon-polaritons in PhCs fabricated in thin slabs of lithium niobate (LN, uniaxial medium) and lithium tantalate (LT, nearly-isotropic medium) and found that in the case of LN the lowest bandgap along the $\Gamma$-$M$ direction of a square lattice forms at a different wavevector than the boundary of the conventional BZ. This experiment will be described in Sec. \ref{sec:exp} following a brief background discussion of the relationship of the BZ to Bragg planes in Sec. \ref{sec:bkgd}.  In Sec. \ref{sec:analys}, we show that the observed atypical bandgap formation is a direct result of the refractive index anisotropy, which distorts the Bragg planes and EM dispersion. Focusing on the case of a two-dimensional (2D) AnPhC with one of the principal axes of refractive index perpendicular to the periodic plane, we construct the Bragg planes BZ (BBZ) using a simple geometrical method involving a WS cell of a distorted reciprocal lattice elaborated in the Appendix.  We demonstrate the usefulness of the BBZ in identifying the symmetry and features of the dispersion bands. In particular, we show that the bandgap between the first and second dispersion bands forms along the BBZ boundary. Furthermore, we apply a more general method of constructing the BBZ to an AnPhC in a thin slab, in which case ``Bragg planes" are generally no longer planes but are represented by curved surfaces. 

\section{\label{sec:bkgd}Background}

In this background section, we briefly review the concept of Bragg planes and the BZ in an IsPhC. Bragg planes are defined in the limit of vanishing periodic perturbation, using the conservation of energy and discrete translational symmetry; the latter conserves the wave vector $\mathbf{k}$  up to a reciprocal lattice vector $\mathbf{G}$. The resulting equations can be summarized as 
\begin{eqnarray}
\omega(\mathbf{k'})&=&\omega(\mathbf{k})\label{eq:1},\\
\mathbf{k'}&=& \mathbf{k+G}\label{eq:2},
\end{eqnarray}
where $\mathbf{k}$ is the incident wavevector and $\mathbf{k'}$  is the diffracted wavevector. For an isotropic medium, we also make use of the dispersion relation 
\begin{eqnarray}
\omega(\mathbf{k})=vk\label{eq:3},
\end{eqnarray}
where $v$  is the speed of light in the medium and  $k$ is the wavevector magnitude.  Eqs. ~\eqref{eq:1} and \eqref{eq:3} imply that the magnitudes of the incident and diffracted wavevectors must be equal, 
\begin{eqnarray}
k'=k.\label{eq:4}
\end{eqnarray} 

Together, Eqs.~\eqref{eq:2} and \eqref{eq:4} imply that the Bragg plane bisects -- at right angles --  the reciprocal lattice vector $\mathbf{G}$ that connects reciprocal lattice points. Brillouin zones can then be constructed using the Bragg planes: the first BZ contains the volume enclosed without crossing any Bragg planes, while higher order BZs contain the volume obtained by crossing subsequent planes. As a result, the WS primitive cell of the reciprocal lattice is bounded by the Bragg planes, and the two methods of defining the BZ are equivalent. 

\begin{figure}
\includegraphics{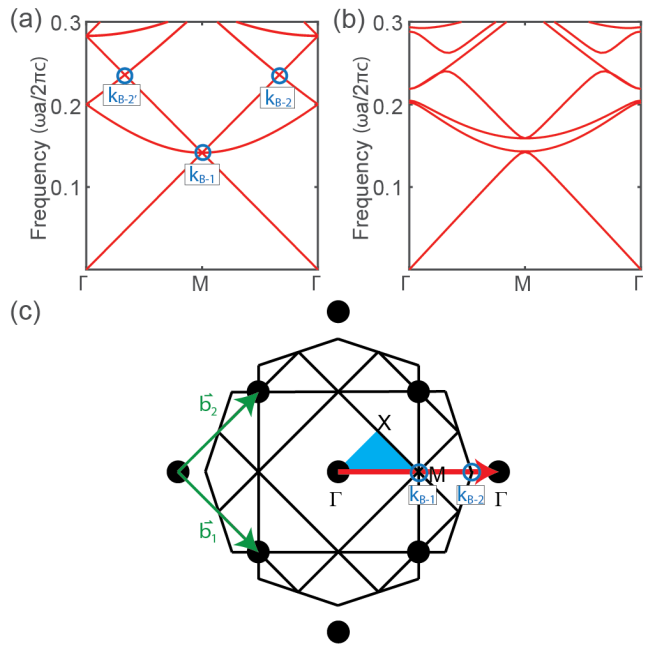}
\caption{\label{fig:1}Crossing points of the empty lattice dispersion curves identify the Bragg planes. (a) TE dispersion curves for propagation along $\Gamma$-$M$-$\Gamma$ in a 2D empty square lattice with an isotropic refractive index $n=5$ and lattice periodicity $a$. The first two crossing points are labelled as $k_{B\text{-}1}$ and $k_{B\text{-}2}$. (b) Corresponding dispersion curves with a finite perturbation from air-holes of radius $r=0.15a$. (c) The first several BZ boundaries in reciprocal space constructed using the WS geometrical method. The propagation direction (red arrow), irreducible Brillouin zone (shaded blue), and reciprocal lattice vectors $b_1$ \& $b_2$ are also indicated. }
\end{figure}

It is instructive to identify the relationship between the Bragg planes and photon dispersion in the limit of vanishing perturbation (i.e index modulation is taken to zero). A dispersion curve with vanishing perturbation is often referred to as an ``empty lattice" or ``free-photon" dispersion curve \cite{Inoue2004,Sakoda2004} in analogy to the free-electron band diagrams in solid state physics \cite{Ashcroft1976a}. In Fig.~\ref{fig:1}a, we show the first few TE bands along $\Gamma$-$M$-$\Gamma$ for an isotropic 2D empty square lattice with refractive index $n=5$ and infinite in the third dimension. Due to periodicity, the empty lattice has folded bands, with crossing points $k_{B\text{-}N}$ (where bands $N$ and $N+1$ cross) corresponding to Bragg planes as defined by Eqs.~(\ref{eq:1}-\ref{eq:2}).  If the perturbation (i.e. periodic index modulation) is finite, crossing points $k_{B\text{-}N}$ not preserved by symmetry will form avoided crossings leading to photonic bandgaps (see Fig.~\ref{fig:1}b). 

In order to demonstrate the correspondence between crossing points $k_B$ and the BZs, in Fig.~\ref{fig:1}c we show the first several BZ boundaries for a square lattice with lattice vectors oriented at 45$^{\circ}$ with respect to the Cartesian axes. The distance from the BZ center $\Gamma$  to the boundaries is equal to the wavevector at which the waves undergo Bragg diffraction. For example, the first BZ boundary encountered along $\Gamma$-$M$-$\Gamma$ is at the high-symmetry $M$ point $k_M=\sqrt{2}\pi/a$; in Fig.~\ref{fig:1}a, this corresponds to the crossing of modes 1 and 2 at $k_{B\text{-}1}$. The second boundary is encountered at an increasing distance from $\Gamma$ and corresponds to the next crossing at $k_{B\text{-}2}$. Crossing points are not unique because the wavevector is only conserved up to the reciprocal lattice vector, $\mathbf{G}$ . For instance, due to discrete translational and time-reversal symmetry,  $k_{B\text{-}2}$ is equivalent to $k_{B\text{-}2'}$ in the first BZ; the same correspondence can also be made for higher-order crossings.  The BZs are instrumental in understanding the formation of the dispersion bands of a photonic crystal. In particular, the lowest bandgap along a particular direction is typically encountered at the boundary of the first BZ.

\section{\label{sec:exp}Experiment}
\begin{figure*}
\includegraphics{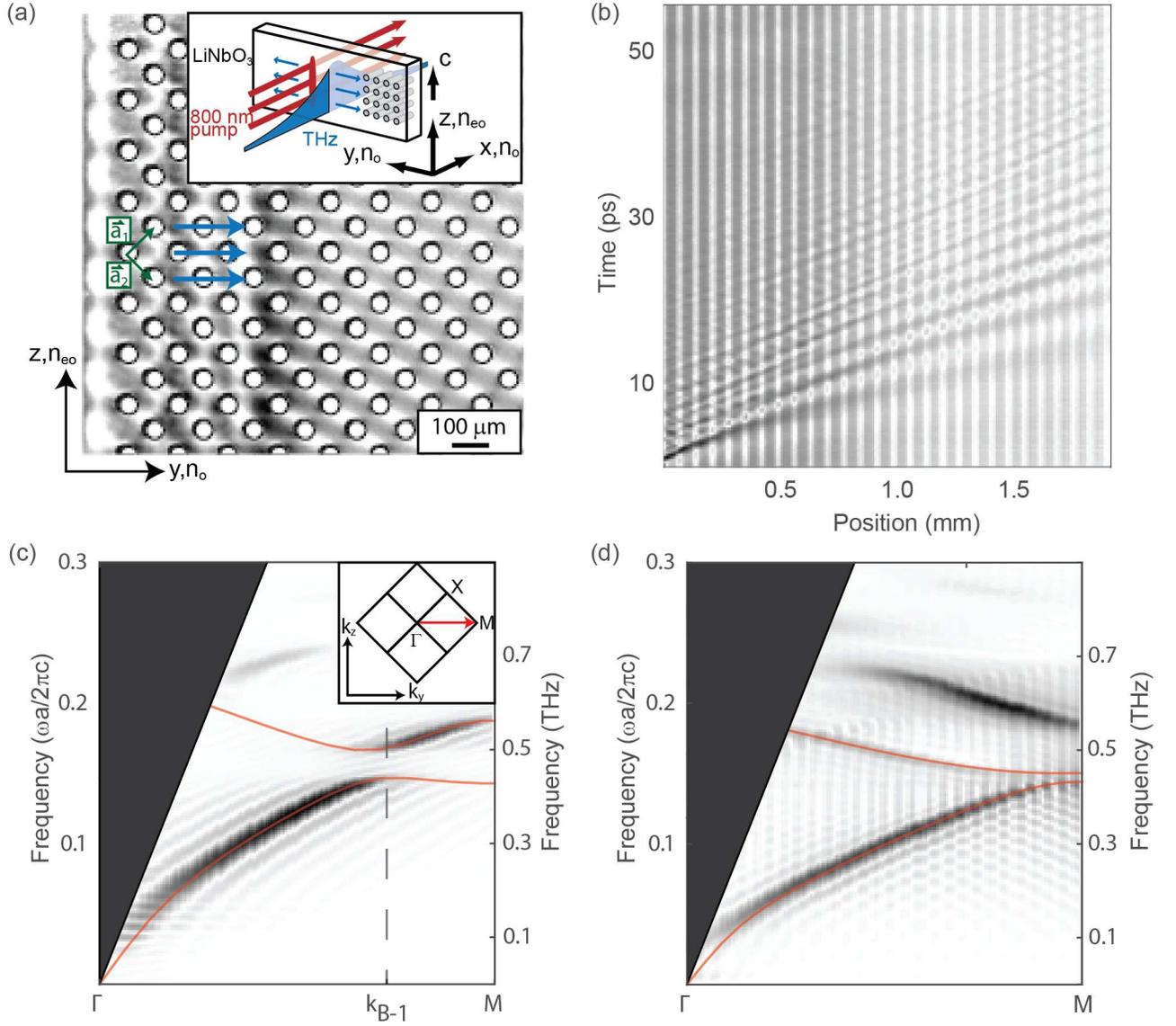}
\caption{\label{fig:2}The experimental observation of atypical bandgap formation. (a) CCD image of a THz wave obtained by using the polaritonics platform, with the image intensity showing the spatially varying THz E-field in the LN PhC slab. The lattice vectors $a_1$ \& $a_2$ are oriented at 45$^{\circ}$ with respect to the principal axes of the LN refractive index. The blue arrows shows propagation along the \textit{y}-axis in real-space. Inset shows the experimental geometry, in which a cylindrically focused 800 nm pump pulse passing through the unpatterned part of the LN slab generates THz waves that counter-propagate along the $\pm$\textit{y}-directions as TE waveguide modes polarized along the \textit{z}-axis. The right-propagating THz wave enters the PhC structure where, at selected times, its E-field spatial distribution is recorded to produce images like the one shown. (b) Space-time plot of the AnPhC in (a), generated by averaging images like the one shown in (a) over the \textit{z}-dimension to produce reduced 1-D images and displaying the images one above another in time order . (c) Polariton dispersion in the LN PhC slab obtained by a 2D Fourier transform of (b), with the location of the Bragg plane at $k_{B\text{-}1}$ shown by a dashed line. Inset shows the BZ with a red arrow indicating the wave vector direction along $k_y$. (d) Polariton dispersion in the isotropic LT PhC slab.  Thin lines in (c) and (d) show calculated dispersion curves for the first two TE-like bands.}
\end{figure*}

We used an optical femtosecond pump-probe technique to excite and dynamically image THz phonon-polaritons in photonic crystal slabs. Our technique permits direct measurement of the dispersion relations inside the PhCs \cite{Ofori-Okai2014a}. The samples were made by using femtosecond laser machining to cut a square lattice of air-holes of radius $r=25.2$ $\mu$m and lattice periodicity $a=100$ $\mu$m  into a 54 $\mu$m thick slab of LN and LT \cite{Sivarajah2013}, where the \textit{c}-axis/extraordinary axis (\textit{eo}-axis) was in the plane of the slab (see Fig.~\ref{fig:2}a \& \ref{fig:2}b inset for orientation). Although LT has a very slight anisotropy in the 0.1-0.6 THz (hereafter simply called ``THz") range that we study ($n_{eo}=6.45$, $n_o=6.46$), it is negligible when compared to the anisotropy of LN ($n_{eo}=5.0$, $n_o=6.7$). Therefore, for the purposes of this discussion we will consider LT isotropic. Measurements were done with the THz polaritonics platform developed for optical generation and imaging of phonon-polaritons in electro-optic crystal slabs \cite{Feurer2003,Feurer2007ARM,Stoyanov2002}. An ultrafast optical pump pulse with 800 nm center wavelength and polarization parallel to the \textit{eo}-axis was cylindrically focused onto the LN slab. The pump pulse generated \textit{eo}-polarized THz-frequency EM waves that were guided along the $\pm$\textit{y}-directions as counter-propagating TE-like modes (\textit{x}-even). Diffraction by the PhC introduces additional wavevector components so that the refractive index anisotropy must be considered. Due to their respective polarizations, waves propagating along the \textit{y}- \& \textit{z}-axes are polarized along the \textit{z}- \& \textit{y}-axes and therefore experience refractive indices of $n_z$ \& $n_y$, respectively. Since the slabs are electro-optic, the propagating THz electric field (E-field) $E_{THz}$ induced a change in the optical-frequency refractive index $\Delta n$ of the slab proportional to the E-field amplitude ($\Delta n\propto E_{THz}$). This index change was detected by using polarization-gating imaging \cite{Werley2013IEEE}, in which a spatially expanded optical probe beam was passed through the slab. The probe beam acquired a phase shift ($\Delta \phi\propto \Delta n \propto E_{THz}$ ) in the presence of the THz E-field, which was converted to an intensity modulation $\Delta I$ by using polarization optics and then detected by a CCD camera. The intensity distribution recorded by the CCD (see Fig 2(a)) corresponds to the THz field pattern in the PhC. 

To extract the dispersion relations of the PhC, we varied the delay of the probe pulse with 200-fs steps and recorded 265 images, capturing the time evolution of the THz field pattern over 54 ps. The recorded   THz field profile was averaged over the \textit{z}-dimension for every time point, yielding a 2D space-time matrix shown in Fig.~\ref{fig:2}b.  Finally, a 2D Fourier transform of the space-time data was used to produce the wave vector-frequency dispersion plot shown in Fig.~\ref{fig:2}c. Many interesting features of LT 2D PhC slabs have been revealed in a previous study using this methodology \cite{Ofori-Okai2014a}. 

Figure~\ref{fig:2}c shows the measured dispersion of the LN PhC slab; the primitive lattice vectors are oriented at 45$^\circ$ with respect to the principal axes of refractive index, such that THz wave propagation along the \textit{y}-direction in real space corresponds to propagation along $k_y$ or $\Gamma$-$M$ in reciprocal space (see Fig.~\ref{fig:2}a).  Overlaid is a calculated dispersion curve of the first two TE-like modes \cite{Joannopoulos2008,Johnson1999a} that shows excellent agreement to the experimental data. For comparison, Fig.~\ref{fig:1}d shows the dispersion curve for the isotropic LT PhC slab with calculated TE-like modes overlaid. The most apparent feature in the LN PhC slab is the location of the first band gap. In the LT PhC, the first bandgap for propagation along $\Gamma$-$M$ is formed, as expected, at the $M$ point.  However, in the LN PhC slab the first bandgap is formed inside the BZ rather than at the BZ boundary. This indicates that the refractive index anisotropy of LN has shifted a Bragg plane from the BZ $M$ point. The experimental dispersion curves provide direct evidence for this shift and thereby demonstrate that -- in an AnPhC -- the WSBZ is no longer bounded by the Bragg planes.  This observation motivated us to construct the BZ of the AnPhC according to the original definition by Brillouin, i.e. by constructing Bragg planes for a lattice in an anisotropic medium.

\section{\label{sec:analys}Analysis}
To begin the analysis, we consider a 2D AnPhC fabricated in a medium with one of the principal axes of refractive index perpendicular to the periodic plane. The AnPhC is assumed to be infinite in the third dimension. Although this is a simpler case than that of a PhC slab, we will see that it yields all the essential features seen in the experiment. We consider the same orientation and index profile as in the experiment, i.e., the PhC is comprised of a square lattice with primitive lattice vectors oriented at 45$^\circ$ with respect to principle axes of refractive index $n_{eo,z}=5.0$ \& $n_{o,y}=6.7$ -- herein referred to as Sqr-45-AnPhC. In this geometry, the modes separate into TE (\textit{x}-even, $E_z$, $E_y$, $H_x$) and TM (\textit{x}-odd, $H_z$, $H_y$, $E_x$). We will consider only TE modes consistent with the experimentally observed TE modes of the PhC slab. We note that for the TM modes the medium is effectively isotropic  \cite{Alagappan2006}. 

\begin{figure}
\includegraphics{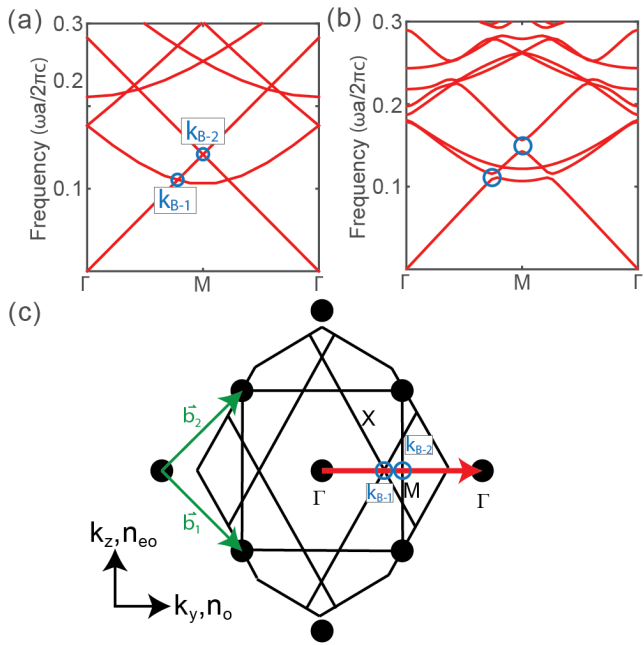}
\caption{\label{fig:3}Crossing points of the empty lattice dispersion curve identify Bragg planes and bandgaps in a 2D AnPhC. (a) Dispersion curves for propagation along $k_y$ in a 2D empty square lattice with lattice periodicity $a$. The first two crossing points (blue circles) are labelled as $k_{B\text{-}1}$ and $k_{B\text{-}2}$. (b) Corresponding dispersion curves with a finite perturbation from air-holes of radius $r=0.15a$. The bandgaps that originate from the crossing points $k_B$ are indicated by blue circles. (c) The first several BBZ boundaries in reciprocal space constructed using Eqs.~(\ref{eq:1}-\ref{eq:2}) and \eqref{eq:6}. The propagation direction $k_y$ (red arrow), reciprocal lattice vectors $b_1$ and $b_2$, and corresponding crossing points are also shown. }
\end{figure}

For the Sqr-45-AnPhC, we first consider propagation along $k_y$ ($\Gamma$-$M$, see Fig.~\ref{fig:1}c), where the first relevant reciprocal lattice vector is $\mathbf{G}=\mathbf{b_1}$  (or equivalently $\mathbf{G}=\mathbf{b_2}$) with
\begin{eqnarray}
\mathbf{b_1}=\frac{2\pi}{a}\left(\frac{1}{\sqrt{2}}\hat{j}-\frac{1}{\sqrt{2}}\hat{k}\right)\label{eq:5}.
\end{eqnarray}
In the anisotropic medium under consideration, the dispersion relation becomes
\begin{eqnarray}
\omega^2(\mathbf{k})=\left(v_{eo}k_y\right)^2+\left(v_{o}k_z\right)^2,\label{eq:6}
\end{eqnarray}
Using Eqs.~(\ref{eq:1}-\ref{eq:2}), (\ref{eq:5}-\ref{eq:6}), and setting $k_z=0$, we can solve for the first crossing point 
\begin{eqnarray}
k_{B-1}=\frac{\sqrt{2}\pi}{a}\frac{v_{eo}^2+v_o^2}{2v_{eo}^2},\label{eq:7}
\end{eqnarray}
where $v_{eo}$ and $v_o$ are the speeds of light along the \textit{y}- and \textit{z}-axes, respectively. This crossing point corresponds to the first Bragg plane encountered along $k_y$; higher order crossing points can be found similarly. In Fig.~\ref{fig:3}a, we plot the corresponding empty lattice dispersion curve along $\Gamma$-$M$-$\Gamma$ where these crossing points can also be observed. In Fig.~\ref{fig:3}b, a finite perturbation from air-holes of radius $r=0.15a$ is introduced to the empty lattice to show the transformation of the crossing points into bandgaps. From these results, it is clear that the wavevector $k_{B\text{-}1}$ in the experimental data of Fig.~\ref{fig:2}c corresponds to a Bragg plane inside the WSBZ \footnote{An infinite 2D PhC is not an exact model for the PhC slab; however the fundamental mode of the slab waveguide in Fig.~\ref{fig:2}c becomes close to a bulk wave for wavevector larger than about  15 rad/mm ($\pi/2a$ for $a=100$ $\mu$m). Therefore the 2D PhC model yields a qualitatively correct model for the slab LN PhC used in the experiment.}.  An equivalent explanation is that the second bands in Fig.~\ref{fig:2}c \& \ref{fig:3} have a polarization component along the ordinary axis. Since $n_o>n_{eo}$ and an environment with a higher refractive index lowers the frequency of the band, the second band is lowered with respective to its isotropic counterpart (e.g. Fig.~\ref{fig:1}a \& \ref{fig:1}b) and crosses the first band inside the WSBZ. 

Figure~\ref{fig:3}c shows Bragg planes constructed using Eqs.~(\ref{eq:1}-\ref{eq:2}) and \eqref{eq:6} in the reciprocal lattice. The domain around the $\Gamma$ point bounded by the Bragg planes corresponds to the BZ per the original definition by Brillouin; we will refer to it as the Bragg-plane Brillouin zone (BBZ).  Just as the WSBZ, the BBZ is a primitive cell of the reciprocal lattice. However, the BBZ tells us more about dispersion in the AnPhC because it identifies the location of Bragg planes where the bandgaps begin to form. In the Appendix, we demonstrate a convenient geometrical method of constructing the BBZ: the reciprocal space is stretched or compressed to make the dispersion relation in the medium look isotropic in the transformed coordinates \footnote{For the TE ($E_x$,$E_y$,$H_z$) waves where the anisotropy is in the \textit{x}-\textit{y} plane, the stretching and compressing of the original coordinate space corresponds to a transformation of Maxwell’s equations to those for an isotropic medium. }. Next we construct the WS cell of the deformed reciprocal lattice, and then deform the lattice back to its original form. 

\begin{figure}
\includegraphics{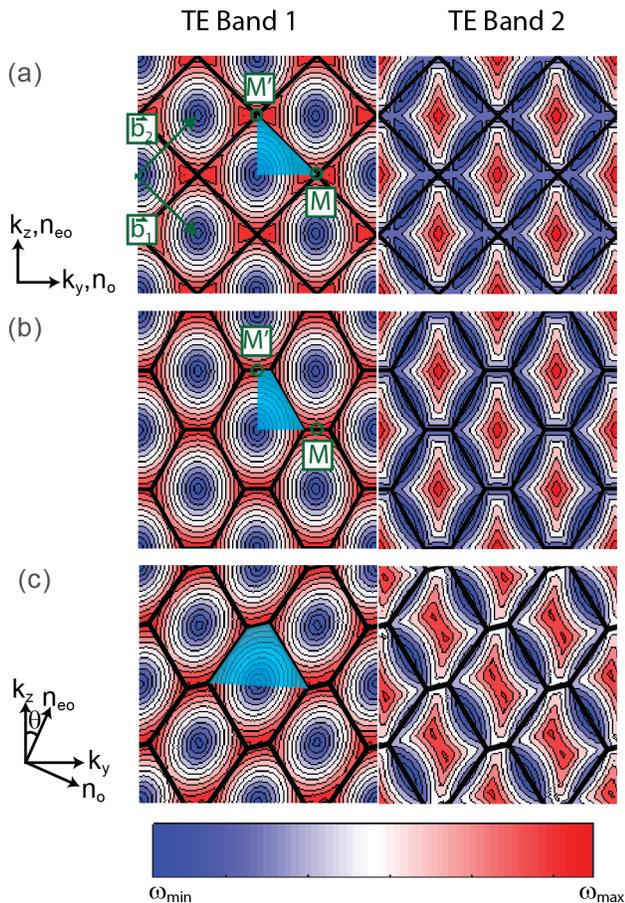}
\caption{\label{fig:4}Isofrequency contour plots demonstrate that the BBZ primitive cell closely reflects the symmetry of EM mode dispersion in an AnPhC. (a) First two numerically solved TE mode IFCs for the Sqr-45-AnPhC with air-hole radius $r=0.1a$. The WSBZ primitive cell and reciprocal lattice vectors $b_1$ \& $b_2$ are also shown. (b)  Same as (a), but with the BBZ primitive cell shown.  (c) The principle axes of refractive index are oriented at $\theta=24^\circ$ with respect to the reciprocal lattice vectors $b_1$ \& $b_2$, which alters the BBZ primitive cell. The associated IBZ (shaded blue) is also shown in all three cases.}
\end{figure}

Just as in the isotropic case, we can use Bragg planes to construct higher-order BBZs. Expectedly, the boundaries of the BBZs encountered along $k_y$ correspond to the crossing points in Fig.~\ref{fig:3}a. The Bragg planes still bisect reciprocal lattice points, but no longer at right angles. As a result, the BBZ retains boundaries at the high-symmetry points identified by WSBZ, but also possess Bragg planes at shifted locations. The primitive unit cell is accordingly a deformed hexagon in contrast to the square WSBZ (see Fig.~\ref{fig:1}c). To provide further insight, we show in Fig.~\ref{fig:4}a the isofrequency dispersion contours (IFCs) for the 2D Sqr-45-AnPhC with air-hole radius $r=0.1a$ along with the corresponding WSBZ primitive cell overlaid.  For comparison, in Fig.~\ref{fig:4}b we show the same IFC but with the BBZ primitive unit cell overlaid. Since primitive cells in general are not unique, both completely characterize the EM dispersion. However, the BBZ primitive cell more closely reflects the symmetry of the dispersion contours because Bragg diffraction plays a dominant role in the formation of dispersion bands in PhCs. To reinforce this point, in Fig.~\ref{fig:4}c we show another BBZ primitive cell when the principal axes of refractive index are oriented at $\theta=24^\circ$ with respect to the reciprocal lattice vectors; even in this arbitrary orientation, the BBZ can accurately represent the symmetry of the EM dispersion.

Besides identifying the Bragg planes, the BBZ also provides an alternative representation of the irreducible Brillouin zone (IBZ) and high-symmetry points. The IBZ is the BZ reduced by point group symmetry of the PhC, which itself is determined by the joint symmetry group of the primitive cell and dielectric tensor \cite{Alagappan2006,Hergert2003,Khromova2008}. The IBZ provides a complete set of non-redundant EM modes within the first BZ. For the Sqr-45-AnPhC in Fig.~\ref{fig:4}a, we show the IBZ constructed using the $C_{2V}$ joint point group symmetry of the WSBZ square primitive cell and the dielectric tensor. In Fig.~\ref{fig:4}b, we show the IBZ constructed using the $C_{2V}$ joint point group symmetry of the BBZ irregular hexagonal primitive cell and the dielectric tensor. For the arbitrary orientation in Fig.~\ref{fig:4}c, the $C_2$ joint point group symmetry means that only half the primitive cell is non-redundant, as shown by its IBZ. Although the IBZs of Fig.~\ref{fig:4}a \& \ref{fig:4}b both represent the non-redundant EM modes, the two offer a different interpretation of the high-symmetry points. The WSBZ shows the high-symmetry points labelled $M$ and $M'$ as non-redundant (i.e. inequivalent) in its IBZ; this conclusion was also drawn by other authors \cite{Alagappan2006,Khromova2008}. However, the BBZ shows $M$ and $M'$ can be defined as equivalent points belonging to different unit cells of the reciprocal lattice, in analogy to the $M$ point in a conventional hexagonal primitive cell BZ.  This behavior is also reflected by the similarity of the dispersion extrema for propagation along $\Gamma$-$M$ in the Sqr-45-AnPhC (see Fig.~\ref{fig:2}c) and along $\Gamma$-$K$-$M$ in the isotropic hexagonal-lattice PhC \cite{Joannopoulos2008}; two Bragg planes are encountered in both cases.

\begin{figure}
\includegraphics{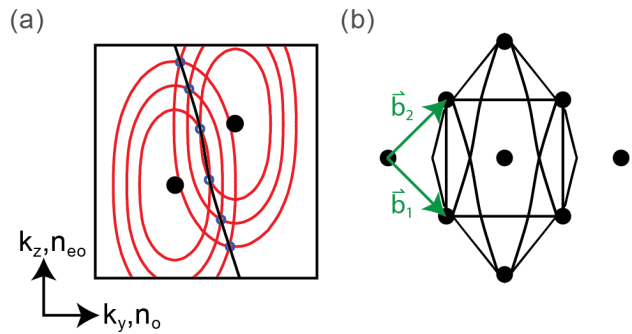}
\caption{\label{fig:5}An AnPhC slab with curved Bragg surfaces. (a) TE-like (\textit{x}-even) isofrequency surfaces at $\omega a/2\pi c=1.32$,1.65,  and 1.98  for an empty square lattice in a $0.54a$ thick slab. The intersection points of the isofrequency surfaces (blue circles) form the Bragg surface (black line) satisfying Eqs.~(\ref{eq:1}-\ref{eq:2}). (b) The first several curved BBZ boundaries in reciprocal space, constructed using the method illustrated in (a).}
\end{figure}

Let us now consider a more complicated case of a thin PhC slab. For a slab with a mirror symmetry such as those in Fig.~\ref{fig:2}, we can classify the modes as TE-like (\textit{x}-even) and TM-like (\textit{x}-odd) \cite{Joannopoulos2008}. In this case, Eq.~\
\eqref{eq:6} is no longer valid due to the dispersion in the slab waveguide. However, Bragg planes can still be identified by solving for the intersection points of the IFCs as specified by Eqs.~(\ref{eq:1}-\ref{eq:2}). This strategy is applied in Fig.~\ref{fig:5} to build the BBZ for the TE-like modes of a Sqr-45-AnPhC slab. The slab has a large anisotropy ($n_{eo,z}=4.5$ \& $n_{o,y}=8.0$) exaggerating the exotic feature of curved Bragg surfaces. A similar curvature also exists in the BBZ of the LN PhC slab used in the experiment, albeit not visually noticeable.  Curvature in the Bragg planes is induced by the combined effect of (i) anisotropy of the underlying slab and (ii) phase velocity dispersion due to waveguide confinement. Alone, neither effect is sufficient to produce curvature as evidenced by 2D AnPhCs and IsPhC slabs: 2D AnPhCs have elliptical isofrequency surfaces but no dispersion, and so their intersections produce flat Bragg planes (e.g. Fig.~\ref{fig:3}c); IsPhC slabs yield a phase velocity dispersion, but the isotropic refractive index results in circular isofrequency surfaces, which also produce flat Bragg planes.

We make a final note here about cases in which TE \& TM waves cannot be decoupled. This occurs generally for 3D AnPhCs as well as for 2D AnPhCs when none of the principle axes of refractive index are perpendicular to the periodic plane. In these instances, the interpretation of the dispersion curves is more complicated due to the hybridizations of the Bloch waves with different polarization. In a PhC slab, accounting for higher-order modes of the waveguide will complicate the situation in a similar way.  One can still construct “Bragg planes” according to Eqs.~(\ref{eq:1}-\ref{eq:2}), but there will be many more of them because  $\mathbf{k}$ and  $\mathbf{k'}$ can now belong to different modes. It is unclear whether the described approach will remain instructive in these situations.  

\section{\label{sec:conc}Conclusion}
We have demonstrated both experimentally and theoretically that for PhCs in optically anisotropic media, the Wigner-Seitz bisection method no longer identifies Bragg planes; consequently the WS cell of the reciprocal lattice is no longer bounded by Bragg planes and thus does not conform to the original definition of the Brillouin zone.  In experiment, we have seen that the lowest bandgap of a PhC fabricated in a LN slab no longer forms at the boundary of the WS cell in the reciprocal space. We have shown that it is possible -- in 2D AnPhCs, as well as in AnPhCs slabs -- to construct an alternative Brillouin zone bounded by Bragg planes, which correctly identifies the location of the experimentally observed bandgap. Furthermore, we have shown that in the presence of dispersion in the underlying material or in a slab waveguide, Bragg ``planes" are generally no longer planes but curved surfaces. Although we have focused on the specific example of a square-lattice PhC with in-plane anisotropy, the conclusions drawn are generalizable to other systems where the modes can be decoupled into TE(-like) and TM(-like). 

For practical tasks such as the calculation of PhC dispersion and eigenmodes, a traditionally defined Wigner-Seitz BZ will still work well, as will, in principle, any primitive cell of the reciprocal lattice. However, we believe that the concept of the alternative Bragg plane BZ will be instrumental in understanding the formation of dispersion bands in AnPhCs. The results should also prove instructive in understanding other Bloch wave systems that possess anisotropy, such as phononic crystals in elastically anisotropic media \cite{Laude2015}.  The results may also inspire promising applications: for example, by actively modulating the anisotropy of a PhC, it should be possible to tune the location of the Bragg planes and therefore the frequency of the zero-group velocity mode. Since the zero-group velocity modes enhance the local density of photonic states \cite{Yang2003}, this may be a valuable handle for frequency tuning spontaneous emission via the Purcell effect \cite{MangaRao2007a,Viasnoff-Schwoob2005}. 

\begin{acknowledgments}
The authors would like to thank Professor Steven G. Johnson for helpful discussions. This work was supported as part of the ``Solid State Solar-Thermal Energy Conversion Centre (S3TEC)," an Energy Frontier Research Centre funded by the U.S. Department of Energy, Office of Science, Office of Basic Energy Sciences under Grant No. DE-SC0001299/DE-FG02-09ER46577 (theoretical modeling) and by the National Science Foundation Grant No. CHE-1111557 (experimental studies). P. Sivarajah acknowledges a Postgraduate Scholarship (PGS-D) from the Natural Sciences and Engineering Research Council of Canada (NSERC). 
\end{acknowledgments}

\appendix*
\section{Wigner-Seitz Construction Of Anisotropic Brillouin Zones Via Coordinate Transformations}
\begin{figure}
\includegraphics{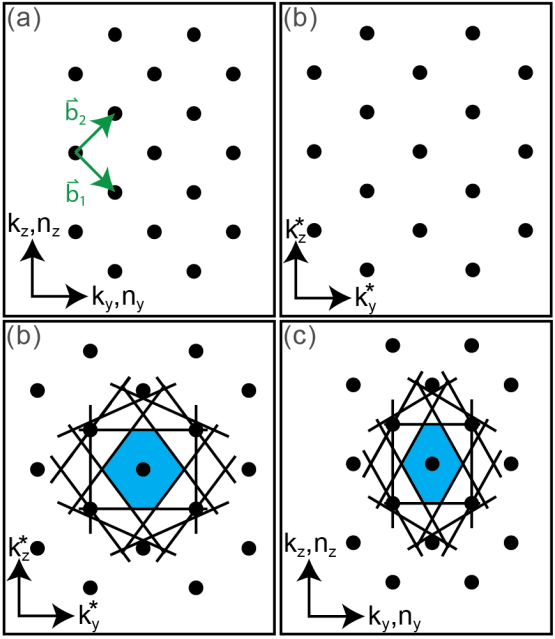}
\caption{\label{fig:A1}Constructing the BBZ of a 2D AnPhC: (a) the original reciprocal lattice; (b) the reciprocal lattice is stretched according to Eqs.~(\ref{eq:A1}-\ref{eq:A2}) and (c) the WS cell of the deformed lattice is constructed; (d) the reciprocal space is deformed back yielding the BBZ of the original reciprocal lattice.}
\end{figure}

We begin by noting that for a 2D AnPhC lattice, we can deform the reciprocal space vectors by a coordinate transformation $(k_z,k_y )\rightarrow(k_z^*  ,k_y^*)$
 \begin{eqnarray}
 k_y^*&=&\left(\frac{v_y}{v_z}\right)k_y,\label{eq:A1}\\
k_z^*&=&k_z.\label{eq:A2}
 \end{eqnarray}
 This transformation results in the isotropic dispersion relation in the transformed reciprocal space, yielding equal magnitudes of the incident and Bragg-diffracted wave vectors, 
 \begin{eqnarray}
 k^{*}{'}=k^*\label{eq:A3}.
 \end{eqnarray}
 This is equivalent to Eq.~\eqref{eq:4} for the isotropic case, where the Bragg plane bisects  $\mathbf{G}$ at right angles, but now in the deformed reciprocal space. In effect, the coordinate transformation transfers the anisotropy from the refractive indices onto the lattice. Thus we construct the BBZ in the transformed reciprocal space as per the WS method, and then transform back to the original coordinate frame to retrieve the correct BBZ. This procedure yields exactly the same BBZ as obtained by analytically solving Eqs.~(\ref{eq:1}-\ref{eq:2}) and \eqref{eq:6}.  In Fig.~\ref{fig:A1}, this geometrical method of constructing the BBZ is illustrated step-by-step for the Sqr-45-AnPhC; the construction yields the BBZ shown in Fig.~\ref{fig:3}b. 
 
\end{document}